\documentclass[12pt,a4paper]{article}
\usepackage{amsmath}

\input{tcilatex}
\begin{document}

\begin{center}
\textbf{Log-Lindley generated family of distributions}\bigskip

\textbf{Lazhar Benkhelifa}\medskip \medskip \medskip

\textit{Laboratory of Applied Mathematics, Mohamed Khider University,
Biskra, 07000, Algeria\medskip\medskip}

\textit{Departement of Mathematics and \textit{I}nformatics, Larbi Ben
M'Hidi University, Oum El Bouaghi, 04000, Algeria\bigskip}

l.benkhelifa@yahoo.fr\textit{\bigskip }\bigskip
\end{center}

\noindent \textbf{Abstract}\smallskip

\noindent A new generator of univariate continuous distributions, with two
additional parameters, called\ the Log-Lindley generated family is
introduced. Some special distributions in the new family are presented. Some
mathematical properties of the new family are studied. The maximum
likelihood method to estimate model parameters is employed. The potentiality
of the new generator is illustrated using a real data set.\bigskip \bigskip

\noindent \textbf{Keywords:} Log-lindley distribution; generated family;
moments; maximum likelihood; order statistics.\bigskip

\noindent AMS Subject Classification: 62E10; 62F03; 62F05; 62F10.\bigskip

\section{\textbf{Introduction}}

In recent years, several authors have proposed various methods for
generating new families of\ univariate continuous distributions, by adding
one or more shape parameters to the baseline distribution, in order to
obtain more flexibility for modeling different types of data sets. For
example, Eugene et al. (2002) proposed the beta generated (G) family of
distributions, Cordeiro and de Castro (2011) proposed the Kumaraswamy-G
family, Alexander et al. (2012) proposed the McDonald-G,
transformed-transformer (T-X) family by Alzaatreh et al. (2013), Weibull-G
by Bourguignon et al. (2014), Lomax-G by Cordeiro et al. (2014), log-gamma-G
by Amini et al. (2014), type-1 half-logistic family of distributions by
Cordeiro et al. (2015), Lindley family of distributions by Cakmakyapan and
Ozel (2016), a new Weibull-G by Tahir et al. (2016), Topp-Leone-G by Rezaei
et al. (2017), Gompertz-G by Alizadeh et al. (2017), odd Lindley-G by
Gomes-Silva et al. (2017).\smallskip \smallskip

\noindent In this paper, we propose a new family of distributions based on
the Log-Lindley distribution. This distribution, with bounded domain,
introduced by Gomez-Deniz et al. (2014) as an alternative to the beta
distribution. They studied the most important properties and gave some nice
applications of this model to insurance and inventory management. The
log-lindley distribution\ is a flexible and simple model obtained by a
logarithmic transformation of the generalized Lindey random variable defined
in Zakerzadeh and Dolati (2009).\medskip

\noindent The probability density function (pdf) and cumulative distribution
function (cdf) are respectively given by%
\begin{equation*}
f\left( x\right) =\frac{a^{2}}{1+ab}x^{a-1}\left( b-\log x\right)
\end{equation*}%
and%
\begin{equation*}
F\left( x\right) =\frac{x^{a}}{1+ab}\left( 1+ab-a\log x\right) ,
\end{equation*}%
for $0<x<1,$ $a>0$ and $b\geq 0.$\medskip

\noindent Motivated by the pioneering work of Cordeiro and de Castro (2011),
we define the Log-Lindley (\textquotedblleft LL for short\textquotedblright
) generator with two extra positive parameters $a$ and $b$ by the following
cdf%
\begin{equation}
F\left( x;a,b,\mathbf{\theta }\right) =\frac{1}{1+ab}\left[ 1+ab-a\log G(x,%
\mathbf{\theta })\right] G^{a}(x,\mathbf{\theta }),
\end{equation}%
where $G(x,\mathbf{\theta })$ is the continuous baseline cdf depending on a
parameter vector $\mathbf{\theta .}$ Then, for an arbitrary parent cdf $G(x)$%
, the LL-G family is defined by the cdf (1). The LL generator has the same
parameters of the baseline $G$ distribution plus two additional parameters $%
a $ and $b$. The density function corresponding to LL-G distribution is%
\begin{equation}
f\left( x;a,b,\mathbf{\theta }\right) =\frac{a^{2}g\left( x,\mathbf{\theta }%
\right) }{1+ab}\left[ b-\log G(x,\mathbf{\theta })\right] G^{a-1}\left( x,%
\mathbf{\theta }\right) ,
\end{equation}%
where $g\left( x,\mathbf{\theta }\right) $ is the baseline pdf. Hereafter, a
random variable $X$ with pdf (2) will be denoted by $X\sim $LL-G$\left( a,b,%
\mathbf{\theta }\right) $. Further, we can omit the dependence on the vector 
$\mathbf{\theta }$ of the parameters and write simply $G(x)=G(x,\mathbf{%
\theta }).$ The hazard rate function of $X$ is given by%
\begin{equation*}
h\left( x\right) =\frac{a^{2}g\left( x,\mathbf{\theta }\right) \left[ b-\log
G(x,\mathbf{\theta })\right] G^{a-1}\left( x,\mathbf{\theta }\right) }{%
1+ab-\left( 1+ab-a\log G(x,\mathbf{\theta })\right) G^{a}\left( x,\mathbf{%
\theta }\right) }.
\end{equation*}%
The quantile function of the LL-G distribution is obtained, by inverting
(1), as%
\begin{equation*}
Q\left( u\right) =Q_{G}\left( \exp \left\{ \frac{1+ab}{a}\right\} \exp
\left\{ \frac{1}{a}W_{-1}\left( -\frac{\left( 1+ab\right) u}{\exp \left(
1+ab\right) }\right) \right\} \right) ,
\end{equation*}%
where $Q_{G}$ is the quantile function of the baseline $G$ distribution and $%
W_{-1}$ is the negative branch of the Lambert $W$ function (i.e., the
solution of the equation $W(z)e^{W(z)}=z$). For more details on Lambert $W$
function, see Section 2 of Jodr\'{a} (2010). If $U$ is a uniform random
variable on the unit interval $\left( 0,1\right) $, then

\begin{equation*}
X=Q_{G}\left( \exp \left\{ \frac{1+ab}{a}\right\} \exp \left\{ \frac{1}{a}%
W_{-1}\left( -\frac{\left( 1+ab\right) U}{\exp \left( 1+ab\right) }\right)
\right\} \right)
\end{equation*}%
will be a LL-G random variable.\bigskip

\noindent The rest of the paper is organized as follows. In Section 2, we
present some new generated distributions. Shapes of the pdf and hazard rate
function are given in Section\ 3. Some mathematical properties of the LL-G
family is derived in Sections 4. Maximum likelihood estimation of the model
parameters and the observed information matrix are presented in Section 5.
In Section 6, an application of the LL-G family is illustrated using a real
data set.. Finally, conclusions are given in Section 7.

\section{\textbf{Special LL-G distributions}}

In this section, we give some special models of the LL-G family of
distributions.

\subsection{\textit{The Log-lindley-normal (LL-N) distribution}}

If $G(x)=\Phi \left( \left( x-\mu \right) /\sigma \right) $ and $g(x)=\phi
\left( \left( x-\mu \right) /\sigma \right) $ are the cdf and pdf of the
normal $N\left( \mu ,\sigma ^{2}\right) $ distribution with parameters $\mu $
and $\sigma ^{2}$, respectively, where $\phi $ and $\Phi $ are the pdf and
cdf of the standard normal distribution, respectively. Then, the LL-N
density function is given by%
\begin{equation*}
f\left( x\right) =\frac{a^{2}\phi \left( \left( x-\mu \right) /\sigma
\right) }{1+ab}\left[ b-\log \left\{ \Phi \left( \left( x-\mu \right)
/\sigma \right) \right\} \right] \left[ \Phi \left( \left( x-\mu \right)
/\sigma \right) \right] ^{a-1}.
\end{equation*}

\subsection{\textit{The Log-lindley-Weibull (LL-W) distribution}}

The pdf of the LL-W distribution is defined from (2) by taking $%
G(x)=1-e^{-\alpha x^{\beta }}$ and $g(x)=\alpha \beta x^{\beta -1}e^{-\alpha
x^{\beta }}$ to be the cdf and pdf of the Weibull distribution respectively.
Then, the LL-W density function is given for $x>0$, by%
\begin{equation}
f\left( x\right) =\frac{a^{2}\alpha \beta x^{\beta -1}e^{-\alpha x^{\beta }}%
}{1+ab}\left[ b-\log \left( 1-e^{-\alpha x^{\beta }}\right) \right] \left[
1-e^{-\alpha x^{\beta }}\right] ^{a-1}.
\end{equation}

\section{\textbf{Shapes}}

We can describe analytically the shapes of the density and hazard rate
functions. The critical points of the pdf of the LL-G distribution are the
roots of the following equation:%
\begin{equation}
\frac{g^{\prime }\left( x\right) }{g\left( x\right) }-\frac{g\left( x\right) 
}{G\left( x\right) \left[ b-\log \left\{ G(x)\right\} \right] }+\left(
a-1\right) \frac{g\left( x\right) }{G\left( x\right) }=0.
\end{equation}%
If $x=x_{0}$ is a root of (4) then it corresponds to a local maximum, a
local minimum or a point of inflexion depending on whether $\lambda \left(
x_{0}\right) <0$, $\lambda \left( x_{0}\right) >0$ or $\lambda \left(
x_{0}\right) =0$, where%
\begin{eqnarray*}
\lambda \left( x\right) &=&\frac{g\left( x\right) g^{\prime \prime }\left(
x\right) -\left[ g^{\prime }\left( x\right) \right] ^{2}}{g^{2}\left(
x\right) }+\left( a-1\right) \frac{g^{\prime }\left( x\right) G\left(
x\right) -g^{2}\left( x\right) }{G^{2}\left( x\right) } \\
&&-\frac{g^{\prime }\left( x\right) G\left( x\right) \left[ b-\log \left\{
G(x)\right\} \right] ^{2}-g^{2}\left( x\right) \left( \left[ b-\log \left\{
G(x)\right\} \right] ^{2}-1\right) }{G^{2}\left( x\right) \left[ b-\log
\left\{ G(x)\right\} \right] ^{3}}.
\end{eqnarray*}%
The critical points of the hazard rate function of the LL-G distribution are
the roots of the following equation:%
\begin{eqnarray}
&&\frac{g^{\prime }\left( x\right) }{g\left( x\right) }-\frac{g\left(
x\right) }{G(x)\left( b-\log \left\{ G(x)\right\} \right) }+\left(
a-1\right) \frac{g\left( x\right) }{G\left( x\right) }  \notag \\
&&-\frac{g(x)G^{a-1}(x)+aG^{a-1}(x)\left[ 1+a\left( b-\log \left\{
G(x)\right\} \right) \right] ^{2}}{\left( 1+ab-\left[ 1+a\left( b-\log
\left\{ G(x)\right\} \right) \right] G^{a}(x)\right) \left( 1+a\left( b-\log
\left\{ G(x)\right\} \right) \right) }=0.\text{ \ \ \ \ \ }
\end{eqnarray}%
If $x=x_{0}$ is a root of (5) then it corresponds to a local maximum, a
local minimum or a point of inflexion depending on whether $\kappa \left(
x_{0}\right) <0$, $\kappa \left( x_{0}\right) >0$ or $\kappa \left(
x_{0}\right) =0$, where%
\begin{eqnarray*}
\kappa \left( x\right) &=&\frac{g^{\prime \prime }\left( x\right) g\left(
x\right) -\left[ g^{\prime }\left( x\right) \right] ^{2}}{g^{2}\left(
x\right) }+\left( a-1\right) \frac{g^{\prime }\left( x\right) G\left(
x\right) -g^{2}\left( x\right) }{G^{2}\left( x\right) } \\
&&-\frac{g^{\prime }\left( x\right) G\left( x\right) \left[ b-\log \left\{
G(x)\right\} \right] -g\left( x\right) \left\{ G\left( x\right) \left[
b-\log \left\{ G(x)\right\} \right] \right\} ^{\prime }}{G^{2}\left(
x\right) \left[ b-\log \left\{ G(x)\right\} \right] ^{2}} \\
&&-\frac{g^{\prime }\left( x\right) G^{a-1}(x)+\left( a-1\right) g^{2}\left(
x\right) G^{a-2}(x)}{\left( 1+ab-\left[ 1+ab-a\log \left\{ G(x)\right\} %
\right] G^{a}(x)\right) \left( 1+a\left( b-\log \left\{ G(x)\right\} \right)
\right) } \\
&&-\frac{ag\left( x\right) G^{a-1}(x)\left[ 2+\left( a-1\right) \left(
1+ab-a\log G(x)\right) G^{-1}(x)\right] }{\left( 1+ab-\left[ 1+ab-a\log
\left\{ G(x)\right\} \right] G^{a}(x)\right) ^{2}\left( 1+ab-a\log \left\{
G(x)\right\} \right) } \\
&&+\frac{g^{2}\left( x\right) G^{-a-2}(x)}{\log \left\{ G(x)\right\} \left(
1+ab-a\log \left\{ G(x)\right\} \right) ^{2}}+\frac{ag\left( x\right)
G^{-a-2}(x)}{\log \left\{ G(x)\right\} } \\
&&+\frac{a^{2}g^{2}\left( x\right) G^{2a-2}(x)\left( b-\log G(x)\right) }{%
\left( 1+ab-\left[ 1+ab-a\log \left\{ G(x)\right\} \right] G^{a}(x)\right)
^{2}\left( 1+ab-a\log \left\{ G(x)\right\} \right) } \\
&&+\frac{a^{3}g\left( x\right) G^{2a-2}\left( 1+ab-a\log G(x)\right) \left(
b-\log G(x)\right) }{\left( 1+ab-\left[ 1+ab-a\log \left\{ G(x)\right\} %
\right] G^{a}(x)\right) ^{2}}.
\end{eqnarray*}

\section{\textbf{Mathematical properties}}

In this section, we present some properties of the LL-G distribution.

\noindent If $\left\vert z\right\vert <1$ and $\delta >0$ is a real
non-integer, we have the following expansions:%
\begin{equation*}
\left( 1-z\right) ^{\delta }=\sum_{j=0}^{\infty }\left( -1\right) ^{j}\binom{%
\delta }{j}z^{j},
\end{equation*}%
and%
\begin{equation*}
\log \left( 1-z\right) =-\sum_{j=0}^{\infty }\frac{z^{j+1}}{j+1}.
\end{equation*}%
By making use the previous expansions, we can obtain%
\begin{equation}
f\left( x\right) =\sum_{i=0}^{\infty }w_{i}h_{i+1}\left( x\right) ,
\end{equation}%
where%
\begin{equation*}
w_{i}=\sum_{k=i}^{\infty }\frac{a^{2}\left( -1\right) ^{k+i}}{\left(
1+ab\right) \left( i+1\right) }\binom{k}{i}\left[ b\binom{a-1}{k}%
+\sum_{j=0}^{\infty }\sum_{l=0}^{j+1}\frac{\left( -1\right) ^{l}}{j+1}\binom{%
j+1}{l}\binom{a+l-1}{k}\right] ,
\end{equation*}%
and $h_{i+1}\left( x\right) =(i+1)g(x)G^{i}(x)$ is the pdf of the
exponentiated-G (exp-G) distribution with power parameter $i+1$. Equation
(6) indicates that the LL-G density function is a linear combination of
exp-G density functions. Therefore, some mathematical properties of the new
model can be derived from those exp-G properties. For example, the \
ordinary and incomplete moments and moment-generating function.

\noindent The $r$th moment of $X\sim $LL-G$\left( a,b,\mathbf{\theta }%
\right) $, is%
\begin{equation*}
E\left( X^{r}\right) =\sum_{i=0}^{\infty }w_{i}E\left( Y_{i+1}^{r}\right) .
\end{equation*}%
where $Y_{i+1}\sim $exp-G$(i+1).\ $Nadarajah and Kotz (2006) gave the
closed-form expressions for moments of several exp-G distributions.\bigskip

\noindent The moment generating function of $X$, say $M(t)=E\left(
e^{tX}\right) $, is given by%
\begin{equation*}
M(t)=\sum_{i=0}^{\infty }w_{i}E\left( e^{tY_{i+1}}\right) .
\end{equation*}

\noindent The order statistics play an important role in reliability and
life testing. Let $X_{1,n},\ldots ,X_{n,n}$ denote the order statistics
obtained from this sample. The pdf of $X_{k,n}$ is given by%
\begin{equation*}
f_{k,n}\left( x\right) =\frac{n!}{\left( n-k\right) !\left( k-1\right) !}%
f\left( x\right) \left[ F\left( x\right) \right] ^{k-1}\left[ 1-F\left(
x\right) \right] ^{n-k},\text{ for }k=1,\ldots ,n.
\end{equation*}%
Since $0<F\left( x\right) <1$ for $x>0$, by using the expansion (9) of $%
\left[ 1-F\left( x\right) \right] ^{n-k}$, we obtain%
\begin{equation*}
f_{k,n}\left( x\right) =\frac{n!}{\left( n-k\right) !\left( k-1\right) !}%
\sum_{j=0}^{n-k}\binom{n-k}{j}\left( -1\right) ^{j}f\left( x\right) \left[
F\left( x\right) \right] ^{k+j-1}.
\end{equation*}

\section{\textbf{Maximum likelihood estimation}}

\noindent Let $x_{1},\ldots ,x_{n}$ be a random sample of size $n$ from LL-G
distributions with unknown $r\times \ 1$ parameter vector $\mathbf{\xi =}%
\left( a,b,\mathbf{\theta }\right) ^{T}.$ The likelihood function is%
\begin{equation*}
L\left( \mathbf{\xi }\right) =\left( \frac{a^{2}}{1+ab}\right)
^{n}\dprod\limits_{i=1}^{n}g\left( x_{i}\right) \left[ b-\log G(x_{i})\right]
G^{a-1}\left( x_{i}\right) .
\end{equation*}%
Then, we obtain the log-likelihood function $\ell \left( \mathbf{\xi }%
\right) $%
\begin{eqnarray*}
\ell \left( \mathbf{\xi }\right) &=&2n\log \left( a\right) +n\log \left(
1+ab\right) +\sum_{i=1}^{n}\log g\left( x_{i};\mathbf{\theta }\right) \\
&&+\left( a-1\right) \sum_{i=1}^{n}\log G\left( x_{i};\mathbf{\theta }%
\right) +\sum_{i=1}^{n}\log \left[ b-\log G(x_{i};\mathbf{\theta })\right] .
\end{eqnarray*}%
The components of the score vector are given as follows%
\begin{align*}
U_{a}(\mathbf{\xi })& =\frac{2n}{a}+\frac{nb}{1+ab}+\sum_{i=1}^{n}\log
G\left( x_{i};\mathbf{\theta }\right) , \\
U_{b}(\mathbf{\xi })& =\frac{na}{1+ab}+\sum_{i=1}^{n}\frac{1}{b-\log G(x_{i};%
\mathbf{\theta })},
\end{align*}%
and%
\begin{equation*}
U_{\mathbf{\theta }}(\mathbf{\xi })=\sum_{i=1}^{n}\frac{\overset{\mathbf{.}}{%
g}\left( x_{i};\mathbf{\theta }\right) }{g\left( x_{i};\mathbf{\theta }%
\right) }+\left( a-1\right) \sum_{i=1}^{n}\frac{\overset{\mathbf{.}}{G}%
\left( x_{i};\mathbf{\theta }\right) }{G\left( x_{i};\mathbf{\theta }\right) 
}+\sum_{i=1}^{n}\frac{\overset{\mathbf{.}}{G}\left( x_{i};\mathbf{\theta }%
\right) }{G(x_{i};\mathbf{\theta })\left( b-\log G(x_{i};\mathbf{\theta }%
)\right) },
\end{equation*}%
where $\overset{\mathbf{.}}{g}\left( x_{i};\mathbf{\theta }\right) =\partial
g\left( x_{i};\mathbf{\theta }\right) /\partial \mathbf{\theta }$ and $%
\overset{\mathbf{.}}{G}\left( x_{i};\mathbf{\theta }\right) =\partial
G\left( x_{i};\mathbf{\theta }\right) /\partial \mathbf{\theta }$.

\noindent The MLE $\widehat{\mathbf{\xi }}=\left( \widehat{a},\widehat{b},%
\widehat{\mathbf{\theta }}\right) ^{T}$ of $\mathbf{\xi }=\left( a,b,\mathbf{%
\theta }\right) ^{T}$ is obtained by solving the non-linear likelihood
equations simultaneously $U_{a}(\mathbf{\xi })=0,$ $U_{b}(\mathbf{\xi })=0$
and $U_{\mathbf{\theta }}(\mathbf{\xi })=0$. The observed information
matrix, for making interval inference, is given by%
\begin{equation*}
\left( 
\begin{array}{ccc}
U_{11} & U_{12} & U_{13} \\ 
U_{21} & U_{22} & U_{23} \\ 
U_{31} & U_{32} & U_{33}%
\end{array}%
\right) ,
\end{equation*}%
where%
\begin{eqnarray*}
U_{11} &=&-\frac{2n}{a^{2}}-\frac{nb^{2}}{\left( 1+ab\right) ^{2}},\  \\
U_{12} &=&\frac{n\left( 1+ab\right) -nab}{\left( 1+ab\right) ^{2}},\  \\
U_{13} &=&\sum_{i=1}^{n}\frac{\overset{\mathbf{.}}{G}\left( x_{i};\mathbf{%
\theta }\right) }{G\left( x_{i};\mathbf{\theta }\right) },
\end{eqnarray*}%
\begin{eqnarray*}
U_{22} &=&-\frac{na^{2}}{\left( 1+ab\right) ^{2}}-\sum_{i=1}^{n}\frac{\log
G(x_{i};\mathbf{\theta })}{\left( b-\log G(x_{i};\mathbf{\theta })\right)
^{2}}, \\
U_{23} &=&\sum_{i=1}^{n}\frac{\overset{\mathbf{.}}{G}\left( x_{i};\mathbf{%
\theta }\right) }{G(x_{i};\mathbf{\theta })\left[ b-\log G(x_{i};\mathbf{%
\theta })\right] }, \\
U_{33} &=&\sum_{i=1}^{n}\frac{\overset{\mathbf{..}}{g}\left( x_{i};\mathbf{%
\theta }\right) g\left( x_{i};\mathbf{\theta }\right) -\left[ \overset{%
\mathbf{.}}{g}\left( x_{i};\mathbf{\theta }\right) \right] ^{2}}{g^{2}\left(
x_{i};\mathbf{\theta }\right) } \\
&&+\left( a-1\right) \sum_{i=1}^{n}\frac{\overset{\mathbf{..}}{G}\left(
x_{i};\mathbf{\theta }\right) G\left( x_{i};\mathbf{\theta }\right) -\left[ 
\overset{\mathbf{.}}{G}\left( x_{i};\mathbf{\theta }\right) \right] ^{2}}{%
G^{2}\left( x_{i};\mathbf{\theta }\right) } \\
&&+\sum_{i=1}^{n}\frac{\overset{\mathbf{..}}{G}\left( x_{i};\mathbf{\theta }%
\right) G(x_{i};\mathbf{\theta })\left( b-\log G(x_{i};\mathbf{\theta }%
)\right) }{\left[ G(x_{i};\mathbf{\theta })\left( b-\log G(x_{i};\mathbf{%
\theta })\right) \right] ^{2}} \\
&&-\sum_{i=1}^{n}\frac{\left[ \overset{.}{G}\left( x_{i};\mathbf{\theta }%
\right) \right] ^{2}\left[ b-1-\log G(x_{i};\mathbf{\theta })\right] }{\left[
G(x_{i};\mathbf{\theta })\left( b-\log G(x_{i};\mathbf{\theta })\right) %
\right] ^{2}},
\end{eqnarray*}%
where $\overset{\mathbf{..}}{g}\left( x_{i};\mathbf{\theta }\right)
=\partial ^{2}g\left( x_{i};\mathbf{\theta }\right) /\partial \mathbf{\theta 
}^{2}$ and $\overset{\mathbf{..}}{G}\left( x_{i};\mathbf{\theta }\right)
=\partial ^{2}G\left( x_{i};\mathbf{\theta }\right) /\partial \mathbf{\theta 
}^{2}$.

\section{\textbf{A real data application}}

In this section, the performance of the proposed family is illustrated by
considering one real data set. This data set is given by Bjerkedal (1960)
which represents the survival time in days of 72 guinea pigs infected with
virulent tubercle bacilli. The data set is: 12, 15, 22, 24, 24, 32, 32, 33,
34, 38, 38, 43, 44, 48, 52, 53, 54, 54, 55, 56, 57, 58, 58, 59, 60, 60, 60,
60, 61, 62, 63, 65, 65, 67, 68, 70, 70, 72, 73, 75, 76, 76, 81, 83, 84, 85,
87, 91, 95, 96, 98, 99, 109, 110, 121, 127, 129, 131, 143, 146, 146, 175,
175, 211, 233, 258, 258, 263, 297, 341, 341, 376. We compare the fit of the
LL-W distribution defined in (6), with the beta W (BW), Kumaraswamy-W (KW),
McDonald-W (McW), Marshall-Olkin-W (MOW), Weibull-W (WW), Lomax-W (LoW),
Lindley-W (LiW), Topp-Leone-W (TW), Gompertz-W (GW), odd Lindley-W (OLW),
where their pdfs are%
\begin{eqnarray*}
f_{TW}\left( x\right) &=&2ab\alpha \beta x^{\beta -1}e^{-\alpha x^{\beta
}}\left( 1-e^{-\alpha x^{\beta }}\right) ^{ab-1} \\
&&\times \left[ 1-\left( 1-e^{-\alpha x^{\beta }}\right) ^{b}\right] \left[
2-\left( 1-e^{-\alpha x^{\beta }}\right) ^{b}\right] ^{a-1}, \\
f_{GW}\left( x\right) &=&a\alpha \beta x^{\beta -1}e^{\left( 1+b\right)
\alpha x^{\beta }}e^{\frac{a}{b}\left( 1-e^{b\alpha x^{\beta }}\right) }, \\%
[0.15in]
f_{LoW}\left( x\right) &=&ab^{a}\alpha \beta \frac{x^{\beta -1}}{\left(
b+\alpha x^{\beta }\right) ^{a+1}}, \\
f_{LiW}\left( x\right) &=&\frac{a^{2}}{\left( 1+a\right) }\alpha \beta
x^{\beta -1}e^{-\alpha ax^{\beta }}\left( 1+\alpha x^{\beta }\right) , \\
f_{OLW}\left( x\right) &=&\frac{a^{2}}{\left( 1+a\right) }\alpha \beta
x^{\beta -1}e^{2\alpha x^{\beta }}\exp \left( -a\left[ e^{\alpha x^{\beta
}}-1\right] \right) , \\
f_{WW}\left( x\right) &=&\frac{ab\alpha \beta x^{\beta -1}e^{-\alpha
x^{\beta }}}{1-e^{-\alpha x^{\beta }}}\left[ -\log \left( 1-e^{-\alpha
x^{\beta }}\right) \right] ^{b-1} \\
&&\times \exp \left( -a\left[ -\log \left( 1-e^{-\alpha x^{\beta }}\right) %
\right] ^{b}\right) , \\
f_{MOW}\left( x\right) &=&\frac{a\beta \alpha x^{\beta -1}e^{-\alpha
x^{\beta }}}{\left[ 1-\left( 1-a\right) e^{-\alpha x^{\beta }}\right] ^{2}},
\\
f_{McW}\left( x\right) &=&\frac{c}{B\left( a/c,b\right) }\alpha \beta
x^{\beta -1}e^{-\alpha x^{\beta }}\left( 1-e^{-\alpha x^{\beta }}\right)
^{a-1}\left[ 1-\left( 1-e^{-\alpha x^{\beta }}\right) ^{c}\right] ^{b-1}, \\
f_{KW}\left( x\right) &=&ab\alpha \beta x^{\beta -1}e^{-\alpha x^{\beta
}}\left( 1-e^{-\alpha x^{\beta }}\right) ^{a-1}\left[ 1-\left( 1-e^{-\alpha
x^{\beta }}\right) ^{a}\right] ^{b-1}, \\
f_{BW}\left( x\right) &=&\frac{\alpha \beta x^{\beta -1}e^{-\alpha bx^{\beta
}}}{B\left( a,b\right) }\left( 1-e^{-\alpha x^{\beta }}\right) ^{a-1},
\end{eqnarray*}

\noindent We estimate the model parameters of the distributions by the
method of maximum likelihood. Table 1 lists the MLEs and the corresponding
standard errors in parentheses of the model parameters.\smallskip

\noindent To choose the best model, we use the maximized loglikelihood ($%
-2\log \left( L\right) $), Akaike information criterion (AIC), Consistent
Akaike information criterion (CAIC), Bayesian information criterion (BIC)
and Hannan-Quinn information criterion (HQIC):%
\begin{eqnarray*}
\text{AIC} &=&-2\log \left( L\right) +2k,\text{ \ \ \ BIC}=-2\log \left(
L\right) +k\log \left( n\right) \text{,} \\
\text{CAIC} &=&-2\log \left( L\right) +\frac{2kn}{n-k-1}\text{\ \ and \ HQIC}%
=-2\log \left( L\right) +2k\log (\log \left( n\right) ),
\end{eqnarray*}%
where $k$ is the number of the model parameters and $n$ is the sample size.
The better distribution to fit the data corresponds to smaller values of
these criteria. Therefore, from Table 2, we conclude that the LL-W
distribution gives an excellent fit than the other models.\smallskip

\section{\textbf{Conclusions}}

\noindent We have introduced a new generated family of distributions, called
the LL-G family. We have presented some special cases. We have proved that
the LL-G density function is a linear combination of exponentiated
distributions which allow us to obtain its properties such as ordinary and
incomplete moments, moment-generating function, mean deviations, Bonferroni
and Lorenz curves, R\'{e}nyi entropy and order statistics. The LL-G family
parameters are estimated by maximum likelihood. An$\ $example to real data
illustrate the performance of the proposed family.

\bigskip 

\bigskip 

\begin{table}[h] \centering%
\caption{MLEs of the model parameters and the corresponding standard errors
given in parentheses. }%
\begin{tabular}{lccccc}
\hline
model & $\alpha $ & $\beta $ & $a$ & $b$ & $c$ \\ \hline
LLW & $0.230493$ & $0.547514$ & $14.024860$ & $0.047958$ &  \\ 
& $\left( 0.312457\right) $ & $\left( 0.211727\right) $ & $\left(
15.529796\right) $ & $\left( 0.078301\right) $ &  \\ 
TW & $0.49220$ & $0.37141$ & $6.00952$ & $4.46868$ &  \\ 
& $\left( 1.88508\right) $ & $\left( 0.46321\right) $ & $\left(
35.48929\right) $ & $\left( 35.73443\right) $ &  \\ 
GW & $0.0030390$ & $0.8408393$ & $2.4757193$ & $6.9645592$ &  \\ 
& $\left( 0.0020311\right) $ & $\left( 0.1001659\right) $ & $\left(
0.9023052\right) $ & $\left( 0.0907703\right) $ &  \\ 
LoW & $0.0030609$ & $2.0150751$ & $1.2104142$ & $23.8855555$ &  \\ 
& $\left( 0.0044717\right) $ & $\left( 0.4139031\right) $ & $\left(
0.4725888\right) $ & $\left( 0.0090636\right) $ &  \\ 
LiW & $1.187295$ & $0.988639$ & $0.017651$ &  &  \\ 
& $\left( 1.309358\right) $ & $\left( 0.080269\right) $ & $\left(
0.018779\right) $ &  &  \\ 
OLW & $2.6443973$ & $0.1650443$ & $0.0068032$ &  &  \\ 
& $\left( 1.2154923\right) $ & $\left( 0.0426117\right) $ & $\left(
0.0103276\right) $ &  &  \\ 
WW & $0.177315$ & $0.535990$ & $9.582126$ & $1.527191$ &  \\ 
& $\left( 0.036533\right) $ & $\left( 0.103519\right) $ & $\left(
0.050421\right) $ & $\left( 0.862084\right) $ &  \\ 
MOW & $0.563061$ & $0.473979$ & $80.404218$ &  &  \\ 
& $\left( 0.266324\right) $ & $\left( 0.076261\right) $ & $\left(
54.900698\right) $ &  &  \\ 
McW & $0.25188$ & $0.63229$ & $12.93719$ & $0.53561$ & $1.44729$ \\ 
& $\left( 0.39718\right) $ & $\left( 0.39417\right) $ & $\left(
19.05401\right) $ & $\left( 0.65902\right) $ & $\left( 29.69634\right) $ \\ 
KW & $0.951877$ & $0.326516$ & $49.594511$ & $1.518053$ &  \\ 
& $\left( 0.268110\right) $ & $\left( 0.107911\right) $ & $\left(
0.025434\right) $ & $\left( 1.428337\right) $ &  \\ 
BW & $0.23649$ & $0.64716$ & $12.15966$ & $0.51710$ &  \\ 
& $\left( 0.34335\right) $ & $\left( 0.36361\right) $ & $\left(
14.91623\right) $ & $\left( 0.58922\right) $ &  \\ 
Weibull & $0.0028431$ & $1.2587947$ &  &  &  \\ 
& $\left( 0.0020601\right) $ & $\left( 0.1406885\right) $ &  &  &  \\ \hline
\end{tabular}%
\end{table}%

\begin{table}[h] \centering%
\caption{The statistics:  -2log(L), AIC, CAIC, BIC and HQIC.}%
\begin{tabular}{lccccc}
\hline
model & $-2\log \left( L\right) $ & AIC & CAIC & BIC & HQIC \\ \hline
LLW & $\mathbf{779.7472}$ & $\mathbf{787.7472}$ & $\mathbf{788.3442}$ & $%
\mathbf{796.8539}$ & $\mathbf{791.3726}$ \\ 
TW & $780.1929$ & $788.1929$ & $788.79$ & $797.2996$ & $791.8183$ \\ 
GW & $798.7376$ & $806.7376$ & $807.3346$ & $815.8442$ & $810.363$ \\ 
LoW & $783.3026$ & $791.3026$ & $791.8997$ & $800.4093$ & $794.928$ \\ 
LiW & $788.9608$ & $794.9608$ & $795.3138$ & $801.7908$ & $797.6799$ \\ 
OLW & $796.4631$ & $802.4631$ & $802.5624$ & $809.0394$ & $804.9285$ \\ 
WW & $780.3174$ & $788.3174$ & $788.9145$ & $797.4241$ & $791.9428$ \\ 
MOW & $792.0679$ & $798.0679$ & $798.4209$ & $804.8979$ & $800.787$ \\ 
McW & $780.0641$ & $790.0641$ & $790.9732$ & $801.4474$ & $794.5958$ \\ 
KW & $780.2858$ & $788.2858$ & $788.8829$ & $797.3925$ & $791.9112$ \\ 
BW & $780.0632$ & $788.0632$ & $788.6602$ & $797.1698$ & $791.6886$ \\ 
Weibull & $795.6583$ & $799.6583$ & $799.8322$ & $804.2116$ & $801.471$ \\ 
\hline
\end{tabular}%
\end{table}%


\begin{thebibliography}{99}
\bibitem{} Alexander, C., Cordeiro, G.M., Ortega, E.M.M. and Sarabia, J.M.
(2012). Generalized beta generated distributions. Computational Statistics
and Data Analysis, 56, 1880--1897.

\bibitem{} Alizadeh, M., Cordeiro G.M., Bastos L. G., Ghosh, P.I. (2017) The
Gompertz-G family of distributions, Journal of statistical theory and
practice, 11, 179-207.

\bibitem{} Alzaatreh, A., Lee, C. and Famoye, F. (2013). A new method for
generating families of distributions. Metron, 71, 63--79.

\bibitem{} Amini, M., MirMostafaee, S.M.T.K. and Ahmadi, J. (2014).
Log-gamma-generated families of distributions. Statistics 48:913--932.

\bibitem{} Bjerkedal, T. (1960). Acquisition of resistance in guinea pigs
infected with different doses of virulent tubercle bacilli. American Journal
of Hygiene 72, 130--148.

\bibitem{} Bourguignon, M., Silva, R.B. and Cordeiro, G.M. (2014). The
Weibull-G family of probability distributions. J. Data Sci., 12, 53--68.

\bibitem{} Cakmakyapan, S. and Ozel, G. (2016). The Lindley Family of
Distributions: Properties and Applications. Hacettepe Journal of Mathematics
and Statistics.

\bibitem{} Cordeiro, G. M. and de Castro M. (2011). A new family of
generalized distributions. Journal of Statistical Computation and
Simulation, 81, 883--898.

\bibitem{} Cordeiro, G.M., Ortega, E.M.M., Popovic, B.V. and Pescim, R.R.
(2014). The Lomax generator of distributions: Properties, minification
process and regression model, Applied Mathematics and Computation 247,
465--486.

\bibitem{} Eugene N., Lee C., Famoye F. (2002). Beta-normal distribution and
its applications. Comm. Statist. Theory Methods. 31, 497--512.

\bibitem{} G\'{o}mez-D\'{e}niz, E., Sordo, M.A., Calder\'{\i}n-Ojeda, E.,
(2014). The Log--Lindley distribution as an alternative to the beta
regression model with applications in insurance. Insur. Math. Econ. 54,
49--57.

\bibitem{} Gomes-Silva, F. Percontini, A, de Brito, E., Ramos M.W.,
Venancio, R., Cordeiro, G. (2017). The Odd Lindley-G Family of
Distributions, Austrian Journal of Statistics, 46, 65-87.

\bibitem{} Jodr\'{a} P. (2010). Computer generation of random variables with
Lindley or Poisson--Lindley distribution via the Lambert W function. Math.
Comput. Simul. 81, 851--859.

\bibitem{} Rezaei, S., Sadr, B. B., Alizadeh, M., and Nadarajah, S. (2017).
Topp-Leone generated family of distributions: Properties and applications.
Communications in Statistics-Theory and Methods, 46, 2893-2909.

\bibitem{} Tahir, M.H., Zubair, M., Mansoor M., Cordeiro G. M., Alizadeh M.
and Hamedani G.G. (2016). A new Weibull-G family of distributions. Hacettepe
Journal of Mathematics and Statistics, 45, 629-647.

\bibitem{} Zakerzadeh H., Dolati A. (2009). Generalized Lindley
distribution. J. Math. Extension 3, 13--25.
\end{thebibliography}
\end{document}